\documentclass[twocolumn,showpacs,preprintnumbers,amsmath,amssymb]{revtex4}

\usepackage{graphicx}
\usepackage{dcolumn}
\usepackage{bm}

\begin{document}


\title{Fast Condensation in a tunable Backgammon model}

\author{S. L. Narasimhan \\ Solid State Physics Division, Bhabha Atomic Research Center, Mumbai-400085, India}

\date{\today}

\begin{abstract}
We present a Monte Carlo study of the Backgammon model, at zero temperature, in which a departure box is chosen at random with a probability proportional to $(2\omega - 1)k + (1 - \omega)N$, where $k$ is the number of particles in the departure box and $N$ is the total number of particles (equivalently, boxes) in the system. The parameter $\omega \in [0,1]$ tunes the dynamics from being slow ($\omega = 1$) to being fast ($\omega = 0$). This parametrization tacitly assumes a two-box representation for the system at any instant of time and $\omega$ is formally related to the 'memory' parameter of a correlated binary sequence. For $\omega < 1/2$, the system undergoes a fast condensation beyond a certain time that depends on $\omega$ and the system size $N$. This condensation provides an interesting contrast to that studied with Zeta Urn model in that the probability that a box contains $k$ particles evolves differently in the model discussed here. 
\end{abstract}

\maketitle


Backgammon model [1-4] is a simple adaptation of the Ehrenfest's particles-in-boxes model, which demonstrates that the slow evolution of a system towards its ground state could be due to {\it entropy barriers} - sets of equal energy states. Every box in this model is assigned a two-valued energy - namely, minus one if it is empty and zero if it is non-empty - so that the ground state of the system corresponds to having all the particles in one box. The system evolves from an initial zero energy state, corresponding to a "no empty box" configuration, towards its ground state by shifting a randomly chosen particle to a randomly chosen box subject to the standard Metropolis criterion. Since a particle from a more populated box is shifted out with higher probability, accumulation of particles in a box is a rare event, which is what is responsible for the {\it entropy barriers}. A speed-up of the system dynamics can be achieved [5] by randomly choosing a non-empty box from which a particle can be shifted out. At zero temperature, the system ultimately reaches its ground state configuration, and density has no role to play in this condensation process.

On the other hand, if an empty box is assigned zero energy whereas a non-empty box is assigned a negative energy proportional to the logarithm of the number of particles in it, then the system has been shown [6] to have a condensation transition at finite temperature for densities above a critical value. The process of condensation in this model, referred to as the {\it Zeta Urn} model, has been shown [7] to have a scaling behaviour and also an aging dynamics.

In this context, it is of interest to know whether a parameter-driven condensation can be realized at all using the Metropolis dynamics of the Backgammon model. Recognizing the formal analogy between this model and a binary sequence with long-range memory [8], we present Monte Carlo evidence to show that a tunable ('memory') parameter exists, below a critical value of which a condensation transition can be realized at zero temperature. 
This study  provides an interesting contrast to the the density-driven condensation of the Zeta Urn model.   

\noindent {\it Two-box problem} - The Metropolis dynamics of the Backgammon model, at zero temperature, has been recognized [9,10] as equivalent to that of a two-box system consisting of a certain number of particles in one box and the remaining particles in the other box. If, at any point of time, the box containing less number of particles is labelled zero and the other box labelled one, then the particles themselves can be identified by the labels assigned to the boxes in which they are found. The configurational state of this $N$ particles in two-box system can therefore be represented by a sequence of zeros and ones, $\{a_i \mid a_i = 0, 1; i = 1,2,\dots,N\}$. This $N$-tuple corresponds to a vertex of an $N$-dimensional unit hypercube. A random walk on the vertices of this hypercube, in the presence of traps at $(0, 0, \dots ,0)$ and $(1, 1, \dots ,1)$, then clearly represents the stochastic evolution of the two-box system. 

Thus, the Backgammon model may be viewed as a stochastic reduction or {\it coarse-graining} of the number of labels available for a particle to just one label ultimately - {\it i.e.,} it describes a stochastic procedure for {\it coarse-graining} a many-alphabets sequence into a single-alphabet sequence [11], after sampling enormous number of binary sequences en route.

Let there be a total of $N$ particles distributed in two boxes such that $n_0$ of them are found in a box labelled '$0$' and the remaining ($n_1 = N - n_0$) are found in the other box labelled '$1$'. We assume that the box containing less number of particles is always labelled '$0$'. The probability, $p_0$, of {\it choosing a particle} from the box labelled '$0$' is given by ({\it a la} Ritort)
\begin{equation}
p_0 \equiv \frac{n_0}{N}
\end{equation}
which is also the probability of {\it choosing the box} labelled '$0$'. By adding and subtracting $N/2$ in the numerator, we are led to a more general definition, 
\begin{equation}
p_0 = \frac{1}{2}\left( 1 - \mu \frac{\left[ N - 2n_0 \right]}{N} \right)
\end{equation}
where $\mu$ is a tunable parameter [8]. What this parameter stands for becomes clear if we rewrite the above definition in the form,
\begin{equation}
p_0 = \omega \frac{n_0}{N} + (1 - \omega)\frac{n_1}{N}\ 
      ;\quad \omega \equiv (1+\mu )/2 \in [0,1]
\end{equation}
Now, $\omega$ may be interpreted as the {\it weight} associated with box zero. Since $\omega$ lies in the unit interval, the parameter $\mu$ ranges from $-1$ to $+1$. The case $\mu = 1$ corresponds to choosing a box with a probability proportional to the number of particles in it, whereas the case $\mu = -1$ corresponds to choosing a box with a probability proportional to the number of particles in the {\it other} box.

While the choice $\mu = 1$ is known to slow down the system, the choice $\mu = -1$ on the contrary is expected to speed up the system; the parameter $\mu$ is therefore helpful in tuning the evolution of the system from being fast to being logarithmically slow. In particular, the mean relaxation time for the system to reach the ground state from an arbitrary initial state should depend on the parameter $\mu$. This parametric dependence as well as the nature of the transition from slow dynamics to fast dynamics can be studied by computing the Mean First Passage Time (MFPT) [10] for the general case of $p_0$ given by Eq.(2).

Using the same notation as in reference [10],             
%
%
%
%
the MFPT $F_{j,j-1}(2L,\mu)$ for a system, consisting of even number of particles ($N = 2L$) [12],to go from state $j$ to state $j-1$ is given by
\begin{widetext}
\begin{equation}
F_{j,j-1}(2L,\mu) = u_j(2L,\mu) + \sum _{n=j+1}^{L}u_n(2L,\mu)\left( 
                                      \prod _{k=j}^{n-1}u_k(2L,\mu) v_k(2L,\mu) \right)           
\end{equation}
\end{widetext}
where
\begin{eqnarray}
u_k(2L,\mu) &=& 2\left( 1 - \mu \frac{\left[ 2L - 2k \right]}{2L}\right)^{-1} \\
v_k(2L,\mu) &=& \frac{1}{2}\left( 1 + \mu \frac{\left[ 2L - 2k \right]}{2L}\right) 
\end{eqnarray}
We can easily compute them and show that they are monotonically increasing functions of $L-j$ for all non-negative values of $\mu$; {\it i.e.,}$F_{j,j-1}(2L,\mu) > F_{j+1,j}(2L,\mu)$ for all values of $\mu \in [0,1]$. 

In fact, guided by the exact result $F_{1,0}(2L,\mu = 1) = 2^{2L-1} - 1$ for $\mu = 1$, we have plotted in Fig.1 the computed values of $ln(F_{1,0}(2L,\mu)$ as a function of $L$ for various values of $\mu$ in the range $0<\mu \leq 1$. It is clear that $F_{1,0}(2L,\mu) \sim a(\mu)^L$ where the values of $a(\mu)$ for the $\mu$ values considered are mentioned in the figure caption. Moreover, for $\mu = 0$, we have the exact linear behaviour $F_{1,0}(2L,\mu = 0)\propto L$. This implies that $F_{1,0}(2L,\mu)$ predominantly contributes to the system dynamics for $\mu \in [0,1]$. 

However, for negative values of $\mu$, the MFPTs do not have this monotonic behaviour. In Fig.2, we have plotted $F_{j,j-1}(2L,\mu)$ as a function of $j\in [2,64]$, with $L = 64$, for various values of $\mu$. Deviation from monotonicity may be observed even for a very small negative value of $\mu (= - 0.005)$. Starting from the state $j = L$, the system slows down until it reaches the state, $j_c(\mu)\leq 64$, after which its evolution is progressively fast. 

It is interesting to note that $F_{1,0}(2L,\mu)$ has the functional form $e^{1/L}$ for $\mu = -1.0$ (Fig.3). Since the probability of choosing a box is proportional to the number of particles in the other box for the case $\mu = -1.0$, we expect that $F_{1,0}(2L\to \infty, \mu = -1.0) \to 1$, which is what we observe also in Fig.3. For any other negative value of $\mu$, $F_{1,0}(2L,\mu)$ does not have a simple functional form.
%
%
\begin{figure}
\includegraphics[width=3.25in,height=2.25in]{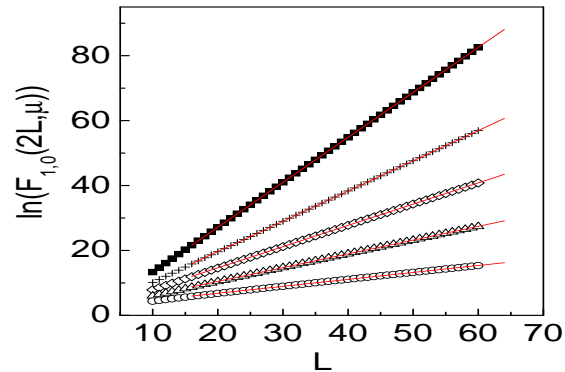}
\caption{Semi-log plot of the MFPT $F_{1,0}(2L,\mu)$ as a function of the system size $L$ for $\mu = 1.0,
0.8, 0.6, 0.4$ and $0.2$ (from top to bottom). The computed data suggest a functional form, $F_{1,0}(2L,\mu) \sim a(\mu)^L$, where $a(\mu)\sim 4, 2.5, 1.9, 1.5$ and $1.2$ respectively for the values of $\mu$
considered. 
}
\end{figure}
%
%
\begin{figure}
\includegraphics[width=3.25in,height=2.25in]{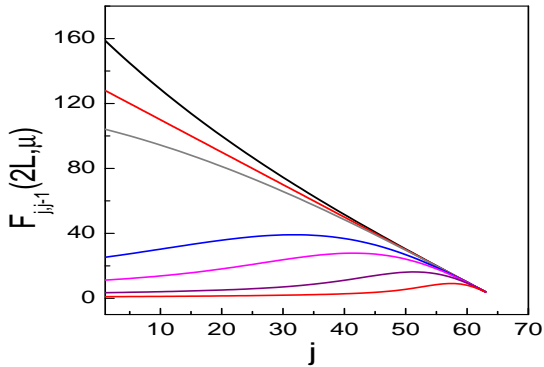}
\caption{A plot of $F_{j,j-1}(2L,\mu)$ as a function of $j\in [2,64]$ for a system of size $L = 64$. The MFPT has been computed for $\mu = 0.005, 0.0, -0.005, -0.1, -0.3$ and $-1.0$ (from top to bottom). }
\end{figure}
%
%
\begin{figure}
\includegraphics[width=3.25in,height=2.25in]{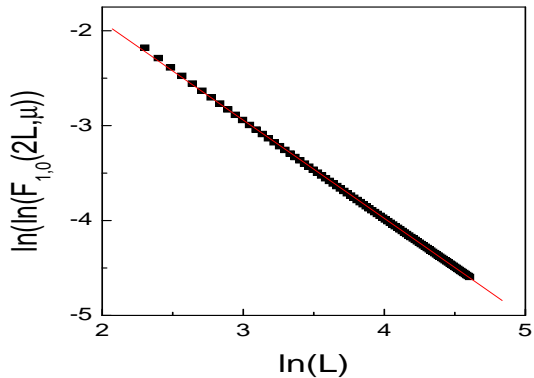}
\caption{The MFPT $F_{1,0}(2L,\mu)$ computed and plotted as a function of the system size $L$ for $\mu = -1.0$. The slope of the straight line is $\sim -1.03$, suggesting thereby the functional form, $F_{1,0}(2L,\mu = -1.0)\sim e^{1/L}$. 
}
\end{figure}

The above MFPT calculations clearly indicate that there is a speed up of the system's evolution for negative values of $\mu$. In order to understand how the more populated boxes gain at the expense of the less populated boxes, we have carried out a Monte Carlo study of this problem.

\noindent {\it Tunable Backgammon model, Monte Carlo study} -  The configurational state of a system of $N$ particles in $N$ boxes at any given time is specified by the $N$-tuple, $\{ n_1, n_2,\dots ,n_N\}$, where $n_k$ is the number of particles found in the $k^{th}$ box. The evolution of the system is based on the dynamical rule that a non-empty box be picked up at random and a particle from it be transferred into one of the remaining non-empty boxes. Thus, at every instant of time, there is a {\it departure-box} and a {\it destination-box}, which may be chosen as follows.

We first assign to each of the non-empty boxes the normalized {\it departure} probability, $d_k \equiv p_k/\sum _k p_k$, where $p_k$ is defined in Eq.(3). We then use these departure probabilities to choose a departure-box at random. On the other hand, a destination-box is just one of the non-empty boxes chosen at random with equal probablity.

Starting from the initial configuration, $\{ n_k = 1\mid k = 1, 2,\dots , N\}$, we obtain the average number of particles per non-empty box, $\lambda (\tau)$, as a function of the Monte Carlo time $\tau$ expressed in units of the system size $N$. In Fig.4, we present the data obtained for a system of size $N = 1024$ corresponding to various values of $\omega$.   
%
%
\begin{figure}
\includegraphics[width=3.25in,height=2.25in]{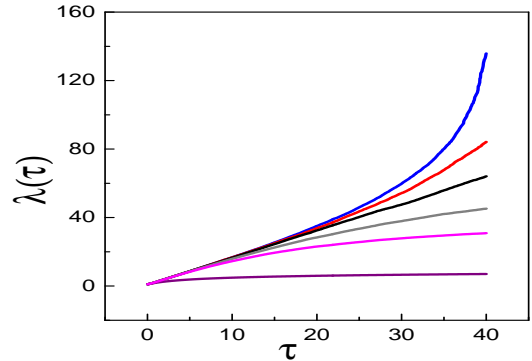}
\caption{Average number of particles per non-empty box, $\lambda(\tau)$, as a function of time. Data correspond to $\omega = 0.0, 0.25, 0.5, 0.75, 0.9$ and $1.0$, from top to bottom, for a system size $N = 1024$. Here, the actual number of Monte Carlo moves is equal to $N\tau$.}
\end{figure}
There is the expected [3] linear growth of $\lambda (\tau)$ with time for $\omega = 1/2$. Linear growth is seen also for smaller values of $\omega < 1/2$ upto a certain time $\tau _0(\omega,N)$ beyond which the growth becomes superlinear ($\tau _0(\omega,N) \to \infty$ as $\omega \to 0.5$). Evidence for the existence of $\tau _0(\omega,N)$ and the subsequent rapid growth is presented in Fig.5. for the case $\omega = 0$. In fact, $\lambda (\tau)$ saturates fast to the value $N$ beyond $\tau _0(\omega = 0,N)$. Quick estimates of $\tau _0(\omega = 0, N)$, presented in the inset as a double logarithmic plot, suggest a power-law dependence $\tau _0(\omega = 0, N)\sim N^{3/2}$.  

On the other hand, for $1/2 < \omega \le 1$, the growth of $\lambda (\tau)$ is linear upto a certain time $\tau _0$ beyond which it becomes sublinear ($\tau _0 \to 0$ as $\omega \to 1$).
It is a well established result [3] that $\lambda (\tau) \propto ln(\tau)$ for the case $\omega = 1$. In order to check whether the asymptotic sublinear growth seen for the other values of $\omega (> 1/2)$ is also of this type, we present $\lambda (\tau)$ as a function of $ln(\tau)$ in Fig.6. 
The simulation data clearly show that the asymptotic growth for $1/2 < \omega < 1$ is also described by the relation, $\lambda (\tau) \propto ln(\tau)$. 

This may be anticipated from the departure probability, $d_k$, assigned to a non-empty box containing $k ( = 1, 2,\dots , N)$ particles:
\begin{equation}
d_k \equiv \frac{p_k}{\sum _k p_k} = \frac{(2\omega - 1)k + (1-\omega)N}
                                          {(2\omega - 1) + (1-\omega)N(1 - f_0)}
\end{equation}
where $f_0$ is the fraction of empty boxes. In the initial stages of the system evolution ($\tau < \tau _0(\omega, N)$, for some $\tau _0$ that depends on $\omega$ and $N$), the number of particles in any given box will be much smaller than the system size ({\it i.e.,} $k << N$). So, whatever be the value of $\omega$, the dominating second term leads to
\begin{equation}
d_k \sim \frac{1}{1-f_0}
\end{equation} 
which is actually the uniform departure probability assigned to the non-empty boxes for the case $\omega = 1/2$.

As the system evolves beyond $\tau _0(\omega, N)$, there will be more boxes containing larger number of particles ({\it i.e.,} $k\sim N$) and therefore, deviations from the behavior dictated by Eq.(8) are expected. That is to say, the first term in Eq.(7) progressively decides the system dynamics. For $\omega > 1/2$, there will be a few highly populated boxes and the positive first term will ensure that the system evolution is similar to what we see for the case $\omega = 1$. On the other hand, for $\omega < 1/2$, the negative first term will ensure that the less populated boxes are emptied fast.

\noindent {\it Summary} - We have shown that the standard Metropolis dynamics of the Backgammon model leads to a condensation process if, at every instant of time, a departure box is chosen at random with a probability given by Eq.(7) for the parameter $\omega < 1/2$. While assigning this parametrized form of departure probability to a non-empty box, we tacitly assume a two-box representation for the system. The cases $\omega = 1/2, 1$ correspond to Model B[3,5] and Model C[1-3] respectively. For all the other values of $\omega$, the system evolves like Model B till a certain time $\tau _0(\omega,N)$ beyond which it behaves like Model C for $\omega > 1/2$ and, on the other hand, undergoes a fast condensation for $\omega < 1/2$. That this condensation provides an interesting contrast to that studied with Zeta Urn model becomes clear if we observe the temporal evolution of the probability that a box contains $k$ particles (Fig. 7, for the case $\omega = 0$). Ultimately, the distribution is made up of two delta functions, one at $k = 0$ and the other at $k = N$.
%
%
\begin{figure}
\includegraphics[width=3.25in,height=2.25in]{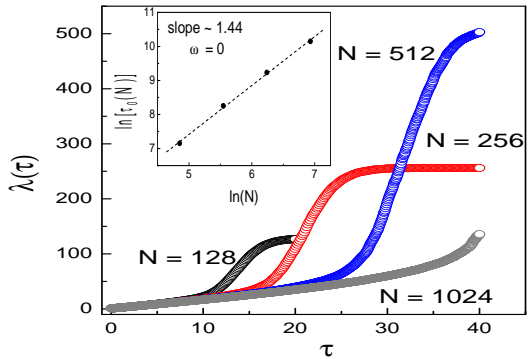}
\caption{Finite size effect on the average number of particles per non-empty box, $\lambda(\tau)$, for the case $\omega = 0$. There is a time, $\tau _0(\omega = 0, N)$, beyond which we see deviation from the linear growth to a rapid saturation at $\lambda(\tau) = N$. Inset: A double logarithmic plot of the approximate estimates of $\tau _0(N)$ for $N = 128, 256, 512$ and $1024$. It suggests a power-law, $\tau _0(N) \sim N^{3/2}$.}
\end{figure}
%
%

%
%
\begin{figure}
\includegraphics[width=3.25in,height=2.25in]{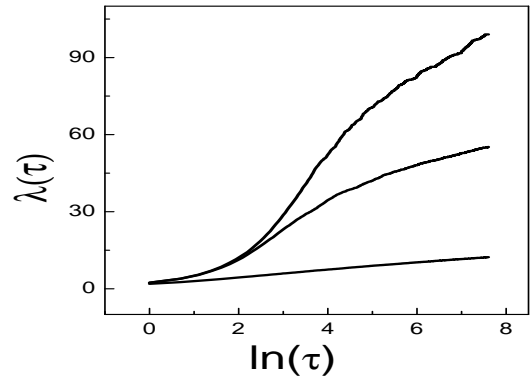}
\caption{$\lambda(\tau)$ as a function of $ln(\tau)$ for $\omega = 0.75, 0.9$ and $1.0$, from top to bottom. System size $N = 1024$.}
\end{figure}
%
%
%
\begin{figure}
\includegraphics[width=3.25in,height=2.25in]{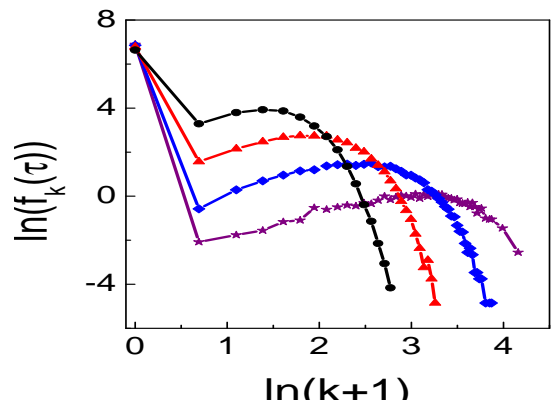}
\caption{Double logarithmic plot of the fraction, $f_k(\tau)$, of boxes having $k$ particles as a function of $k$ at various times $\tau = 8, 16, 32$ and $64$ (top to bottom) for the case $\omega = 0$. System size $N = 1024$.}
\end{figure}

I thank K. P. N. Murthy for helpful discussions.
\begin{enumerate}
\item F. Ritort, Phys. Rev. Lett. {\bf 75}, 1190 (1995)
\item S. Franz and F. Ritort, Europhys. Lett. {\bf 31}, 507 (1995)
\item C. Godreche, J. P. Bouchaud and M. Mezard, J. Phys. A: Math. Gen. {\bf 28} L603 (1995)
\item C. Godreche and J. M. Luck, J. Phys. A: Math. Gen. {\bf 30}, 6245 (1997)
\item B. J. Kim, G. S. Jeon and M. Y. Choi, Phys. Rev. Lett. {\bf 76}, 4648 (1996)
\item P. Bialas, Z. Burda and D Johnson, Nucl. Phys. B {\bf 493}, 505 (1997);
         Bialas, Z. Burda and D Johnson, Nucl. Phys. B {\bf 542}, 413 (1999);
         Bialas, L. Bogacs, Z. Burda and D Johnson, Nucl. Phys. B {\bf 575}, 599 (2000)
\item J. M. Drouffe, C. Godreche and F. Camia, J. Phys. A: Math. Gen. {\bf 31}, L19 (1998);
      C. Godreche and J. M. Luck, {\it cond-mat}/0106272;
      C. Godreche and J. M. Luck, {\it cond-mat}/0109213
\item O. V. Usatenko and V. A. Yampolskii, Phys. Rev. Lett. {\bf 90}, 110601 (2003);
      S. Hod and U. Keshet, Phys. Rev. E {\bf 70}, 015104(R) (2004);
      G. M. Scutz and S. Trimper, Phys. Rev. E {\bf 70}, 045101(R) (2004)
\item A. Lipowsky, J. Phys. A: Math. Gen. {\bf 30}, L91 (1997)
\item K. P. N. Murthy and K. W. Kher, J. Phys. A: Math. Gen. {\bf 30}, 6671 (1997)
\item The statistical analysis of a many-alphabets symbolic sequence such as a natural language text may be simplified by reducing the number of alphabets in the sequence; the motivation here is to filter out the unwanted short-range correlations.
\item Since the even-odd differences become vanishingly small for large values of $N$, we only
consider the case when $N$ is large and even. 
\end{enumerate}  
\end{document}